\documentclass[12pt]{article}

\usepackage{amssymb,amsmath}
\begin{document}

\title{Resonances, Gamow Vectors and Time Asymmetric Quantum Theory\footnote{XXII Symposium on Nuclear Physics, Oaxtepec, Mexico, January (1999).}}

\author{Arno R.~Bohm \and Raymond Scurek \and Sujeewa Wikramasekara \\
\\
Physics Department \\
The University of Texas \\
Austin, Texas 78712 \\
email: bohm@physics.utexas.edu}

\maketitle

\nopagebreak

\begin{abstract}
\footnotesize

\noindent Decaying states can be represented by Gamow vectors with an
exponential, asymmetric time evolution.  This asymmetric evolution is a
manifestation of irreversibility on the microphysical level.  The Rigged
Hilbert Space provides a mathematical theory.

\end{abstract}

\newpage

\section{Introduction: Resonances and Decaying States}

Resonances and quasistable particles play an important role in many areas of physics and in particular in nuclear physics~\cite{1}.  It was in connection with the $\alpha$-decay of a nucleus that the notion of decaying state wave functions was first introduced into quantum mechanics~\cite{2}.  Gamow's wave functions are singular objects like the Dirac kets, and neither Dirac kets $|E,\,l,\,l_3\rangle$ nor Gamow vectors

\begin{equation}
|\psi^G(t)\rangle \,=\, e^{-iHt}|E_R - i\frac{\Gamma}{2},\,l,\,l_3\rangle \label{eq:gamow}
\end{equation}

\noindent are in the Hilbert space $\cal H$.

A Gamow vector is supposed to be an eigenvector of the Hamiltonian $H$ with complex eigenvalue and exponential time evolution

\begin{equation}
H|E_R-i\frac{\Gamma}{2}\rangle \,=\, (E_R-i\frac{\Gamma}{2})|E_R-i\frac{\Gamma}{2}\rangle \label{eq:gamowcomeigen}
\end{equation}

\begin{equation}
e^{-iHt}|E_R-i\frac{\Gamma}{2}\rangle \,=\, e^{-iE_Rt}e^{-\frac{\Gamma}{2}t}|E_R-i\frac{\Gamma}{2}\rangle  \label{eq:gamowev}
\end{equation}

\noindent which would require a Hamiltonian that is not self adjoint (i.e., a ``complex'' Hamiltonian).  In contrast, Dirac kets are considered eigenvectors of $H$ with real eigenvalues, and were, therefore, deemed more acceptable than Gamow vectors.  However, both are mathematically ill defined.  The main reason Gamow vectors are in great disrepute while Dirac kets (scattering state vectors) are copiously used is probably because for Dirac kets the probability density $|\langle\vec{r}|E,\,l,\,l_3\rangle|^2$, though not normalizable ($\int d^3x |\langle\vec{r}|E,\,l,\,l_3\rangle|^2 \sim \infty$), is at least bounded everywhere.  However, for Gamow vectors, not even the probability density $|\langle\vec{r}|\psi^G(t)\rangle|^2 = |\psi^G(r,\theta,\phi,t)|^2$ makes sense.  It is given by

\begin{equation}
|\psi^G(\vec{r},t)|^2 \sim e^{-\Gamma(t-\frac{mr}{|\vec{p}|})}, \,\,\,\, |\vec{p}| = \sqrt{2mE_R}  \label{eq:gamowprobden}
\end{equation}

\noindent which increases exponentially for large $r$ and large negative values of time $t$.  Consequently, it is neither bounded nor normalizable.  Therefore, for Gamow states $\psi^G(t)$ the probability of detection in a finite volume $\Delta V$ at a large distance $r$, $\int_{\Delta V} |\psi^G(\vec{r},t)|^2 d^3\vec{r}$, increases exponentially.

The origin of this ``exponential catastrophe'' (\ref{eq:gamowprobden}) is the assumption that $|\langle\vec{r}|\psi^G(t)\rangle|^2 = |\psi^G(\vec{r},t)|^2$ is the probability density for all time beginning at $t \rightarrow -\infty$.  In reality, the emission of the decay product (e.g., $\alpha$-particle) began at some finite time in the past which we denote by $t_0$ and choose to be $t_0 = 0$.

Since the emission process began at $t_0=0$, and the emitted particle travelled with a speed $v = \frac{p}{m}$, it will reach the region at a distance $r$ from the emitter at the time $t_0(r) = \frac{mr}{p} = \frac{mr}{\sqrt{2mE_R}}$.  Therefore, for $t < t_0(r)$ no decay product can be detected at $r$, i.e., the probability density near $r$ must be zero at times $t < t_0(r)$ (if there are some detector counts, they must be discarded as noise).  It only makes sense to ask what the probability density is for decay products that can be registered near $r$ for times $t \geq t_0(r) > 0$ and which, therefore, were the results of an emission that started at $t_0(0) = t_0 = 0$.  Consequently, the probability density (\ref{eq:gamowprobden}) should be written

\begin{eqnarray}
|\psi^G(\stackrel{\rightarrow}{r},t)|^2 & \sim & e^{-\Gamma(t-t_0(r))} \,\, \mbox{for} \,\, t \geq t_0(r) \\
|\psi^G(\stackrel{\rightarrow}{r},t)|^2 & \sim & 0 \,\, \mbox{for} \,\, t < t_0(r)
\end{eqnarray}

\noindent This means a detector at a distance $r$ detects a counting rate that starts at $t = t_0(r)$ and decreases exponentially with increasing time.  As a consequence, when correctly interpreted, Gamow wave functions predict an exponential decay law and not an ``exponential catastrophe.''

The correct phenomenological interpretation demands that $U^{\dagger}(t) = e^{-iHt}$ in (\ref{eq:gamow}) and (\ref{eq:gamowev}) cannot be the usual unitary-group time evolution with $-\infty < t < \infty$.  Rather, the time evolution of the Gamow kets should be given by a semigroup with $t_0 \leq t < \infty$ (where we choose the physical time $t=t_0$ of emission as the mathematical semigroup time $t=0$, i.e. use the parameter $t' = (t-t_0) \rightarrow t$).  We will denote this semigroup evolution operator by $U_{+}^{\times}(t)$ because it will turn out to be a uniquely defined extension of the unitary-group evolution operator $U^{\dagger}(t)$ for $t \geq 0$, i.e. $U^{\dagger}(t) \subset U_+^{\times}(t)$.

The time evolution of a Gamow vector is then

\begin{equation}
\psi^G(t) \equiv U_+^{\times}(t)\psi^G = e_{+}^{-iH^{\times}t}\psi^G = e^{-iE_Rt}e^{-\frac{\Gamma t}{2}}|E_R-i\frac{\Gamma}{2}\rangle \,\,\, \mbox{for} \,\,\, t \geq 0 \,\,\, \mbox{only}   \label{eq:gamowtimeev}
\end{equation} 

\noindent with $H^{\times} \supset H^{\dagger}$ a uniquely defined extension of the Hilbert space Hamiltonian $H^{\dagger}$.  This means that we have to generalize the unitary-group evolution of standard quantum mechanics with $-\infty < t < \infty$ to a semigroup evolution with $0 \leq t < \infty$.  This also means that for the solutions $\psi^G(t)$ of the time symmetric Schr\"{o}dinger equation we have to choose time {\it asymmetric} boundary conditions, specifically purely outgoing boundary conditions~\cite{3}.

In the Hilbert space formulation of quantum mechanics, the symmetry transformations (e.g., Galilean transformations, Poincare transformations) are described by a unitary group representation in $\cal H$.  Thus, the time evolution is unitary and reversible, and it is given by $U^{\dagger}(t) = e^{-iHt}$, $-\infty < t < \infty$~\cite{4}.  However, there seems to be no empirical reason for disallowing time asymmetric quantum mechanics, characterized by a time evolution semigroup.  Still, the widespread conclusion from this Hilbert space property seems to have been that the irreversible time evolution of isolated quantum mechanical systems (e.g., (\ref{eq:gamowtimeev})) is impossible.

A consequence of this is the pervasive opinion that resonances and decaying states are complicated objects and cannot be represented -- in analogy to the stable states -- by simple, exponentially decaying state vectors like $\psi^G$ in~(\ref{eq:gamowtimeev}).  However, empirical evidence does not suggest that quasi-stable particles are qualitatively different from stable particles.  Stability or the value of lifetime is not a criterion for elementarity.  A particle simply decays if it can and remains stable ($\Gamma = 0$) if selection rules for some quantum numbers prevent it from decaying.  Stable and quasi-stable states should, therefore, be described on the same footing, as, for instance, if both are defined by poles of the analytically continued S-matrix at the pole position $z_R$ or if both are represented as a (generalized) eigenvector of the Hamiltonian with eigenvalue $z_R = E_R-i\frac{\Gamma}{2}$, where $\Gamma = 0$ for stable particles.

State vectors that represent decaying states are used in phenomenological ``effective theories,'' which have been enormously successful.  They describe unstable states as eigenvectors of an ``effective Hamiltonian'' with complex eigenvalues $E_R-i\frac{\Gamma}{2}$, where $E_R$ is the energy of the resonance and $\frac{\hbar}{\Gamma}=\tau$ is the lifetime of the particle.  They describe a time evolution which obeys a simple exponential law.  Examples of such effective theories are the approximate methods of Weisskopf and Wigner and of Heitler for atomic decaying states~\cite{6} and the Lee-Oehme-Yang effective two dimensional theory of the neutral Kaon system~\cite{7}.  In addition, there are many more finite dimensional models of complex, diagonalizable Hamiltonians, in particular in nuclear physics~\cite{1,8}.  Further, finite dimensional models with non-diagonalizable complex Hamiltonian matrices (Jordan blocks) have been considered.  They led to vectors with a non-exponential time evolution~\cite{9}.

In the conventional Hilbert space quantum mechanics ``there does not exist ... a rigorous theory of which these methods can be considered approximations''(M. Levy~\cite{10}).

\section{A Time Asymmetric Quantum Theory in the Rigged Hilbert Space}

The simplest modification of conventional quantum mechanics that allows Hamiltonian generated semigroup evolution is obtained by choosing, instead of the Hilbert space ${\cal H}$, a locally convex space $\Phi$ for physical states and observables.  If one also wants to have the Dirac formalism (i.e. kets, the continuous basis vector expansion, and an algebra of obsevables defined everywhere in $\Phi$), then one has to choose the Rigged Hilbert Space (RHS)~\cite{9a}:

A Rigged Hilbert Space $\Phi \subset {\cal H} \subset \Phi^{\times}$ is a triplet of spaces obtained as three different topological completions of the same algebraic (pre-Hilbert) space $\Phi_{alg}$.  The locally convex, nuclear topology of the space $\Phi$ is given by a countable number of norms, one of which is the usual Hilbert space norm according to which the Hilbert space topology is defined.  The space $\Phi^{\times}$ is the dual of the space $\Phi$, i.e. it is the space of continuous antilinear functionals $|F\rangle$ on $\Phi$:

\begin{equation}
|F\rangle: \,\, \phi \in \Phi \rightarrow F(\phi) \equiv \langle\phi|F\rangle \in \mathbb{C}
\end{equation}

\noindent (According to a theorem by Frechet and Riesz, the Hilbert space $\cal H$ is its own dual ${\cal H} = {\cal H}^{\times}$.)  Thus, the bra-ket $\langle\phi|F\rangle$ is an extension of the scalar product $(\phi|f)$, $f \in {\cal H}$.

The space $\Phi$ contains well-behaved vectors which represent states that can be prepared (created) or registered (detected) by an experimental apparatus.  In contrast, the space $\Phi^{\times}$ contains Dirac kets and Gamow vectors.  The particular choice of $\Phi$ depends on the physical problem at hand, but $\Phi$ is always chosen such that the algebra of observables is represented by an algebra of continuous operators on $\Phi$.

An observable is postulated to be represented by an operator $A$ on $\Phi$, which is continuous with respect to the topology in $\Phi$ (i.e., $\tau_{\Phi}$-continuous),  and by its extensions to $\cal H$ and $\Phi^{\times}$.  Then, corresponding to the triplet of spaces

\begin{equation}
\Phi \subset {\cal H} = {\cal H}^{\times} \subset \Phi^{\times}
\end{equation}

\noindent we have a triplet of operators

\begin{equation}
A^{\dagger}|_{\Phi} \subset A^{\dagger} \subset A^{\times}
\end{equation}

\noindent where $A^{\dagger}$ is the Hilbert space adjoint of $A$, and $A^{\times}$ is the extension of $A^{\dagger}$ to the space $\Phi^{\times}$.  The operator $A^{\times}$ is defined for any $\tau_{\Phi}$-continuous operator $A$ by the identity

\begin{equation}
\langle A\phi|F\rangle = \langle\phi|A^{\times}F\rangle \, \,\,\, \forall \, \phi \in \Phi \, \mbox{and} \, \forall \, F \in \Phi^{\times}
\end{equation}

\noindent If $A$ is a $\tau_{\Phi}$-continuous operator, then $A^{\times}$ is also $\tau_{\Phi^{\times}}$-continuous (but in general not $\tau_{\cal H}$-continuous) and a generalized eigenvector $|F\rangle \in \Phi^{\times}$ of $A$ with eigenvalue $\omega$ is defined as a vector which satisfies

\begin{equation}
\langle A\phi|F\rangle = \langle \phi|A^{\times}F\rangle = \omega\langle\phi|F\rangle \,\,\, \forall \phi \in \Phi
\end{equation}

\noindent Since this is true for all $\phi \in \Phi$, this can be written more abstractly as $A^{\times}|F\rangle = \omega|F\rangle$.

Among the generalized eigenvectors of the Hamiltonian $H$ in $\Phi^{\times}$ are the Dirac kets

\begin{equation}
H^{\times}|E\rangle = E|E\rangle, \,\,\,\, E \geq 0
\end{equation}

\noindent and the Gamow vectors (or Gamow kets)

\begin{equation}
H^{\times}|E_R-i\frac{\Gamma}{2}\rangle = (E_R-i\frac{\Gamma}{2})|E_R-i\frac{\Gamma}{2}\rangle    \label{eq:gamowvector}
\end{equation}

\noindent The Hamiltonian $H$ is always assumed to be (essentially) self-adjoint and bounded from below.  Nevertheless, generalized eigenvectors can have complex eigenvalues.

For a given Hilbert space ${\cal H}$, one can have different (locally convex) spaces $\Phi$.  This allows us to choose different subspaces of $\cal H$ for different physical interpretations.  In particular, we can choose one set of vectors $\phi^+ \in \Phi_-$ to represent states $|\phi^+)(\phi^+|$ or $W = \sum_{i} w_i|\phi_i^+)(\phi_i^+|$ that are prepared by a preparation apparatus (e.g., an accelerator) and a different set of vectors $\psi^- \in \Phi_+$ to represent observables $|\psi^-)(\psi^-|$ or $A = \sum_{i} a_i|\psi_i^-)(\psi_i^-|$, which are measured (or registered) by a registration apparatus (e.g., a detector in a typical scattering experiment).  In Hilbert space quantum mechanics, one assumes $\Phi_- \equiv \Phi_+ \equiv \Phi \subset {\cal H}$ or even $\Phi_{\pm} \equiv {\cal H}$.  One of the fundamental aspects of the new RHS quantum theory is to distinguish meticulously between states (e.g., in-states $\phi^+$ of a scattering experiment) and observables (e.g., so-called out-states or out-observables $\psi^-$ of a scattering experiment).  A justification for this is the fundamental principle that {\it before an observable can be measured in a state, the state must be prepared}.  We call this truism the preparation $\Rightarrow$ registration arrow of time~\cite{11}.  It is a statement of the vague notion of causality and the phenomenological basis of our empirical deduction in section 1 of the semigroup time evolution (\ref{eq:gamowtimeev}).  Therefore, for each quantum physical system we need two RHS's, one for the prepared states $\phi^+$ (which are defined by the preparation apparatus)

\begin{equation}
\phi^+ \in \Phi_- \subset {\cal H} \subset \Phi_-^{\times}  \label{eq:rhsin}
\end{equation}

\noindent and one for the registered observables $\psi^-$ (which are defined by the registration apparatus)

\begin{equation}
\psi^- \in \Phi_+ \subset {\cal H} \subset \Phi_+^{\times}  \label{eq:rhsout}
\end{equation}

\noindent with the same $\cal H$ in (\ref{eq:rhsin}) and (\ref{eq:rhsout}) and for which $\Phi_+ \cap \Phi_- \neq {0}$.

The space $\Phi_-^{\times}$ contains in-states and exponentially growing Gamow vectors, while the space $\Phi_+^{\times}$ contains out-states and exponentially decaying Gamow vectors.

\begin{equation}
|E,\,l,\,l_3\,^{\mp}\rangle \in \Phi_{\pm}^{\times}
\end{equation}

\begin{equation}
|E_R \mp i\frac{\Gamma}{2},\,l,\,l_3\,^{\mp}\rangle \in \Phi_{\pm}^{\times}
\end{equation}

From the preparation $\Rightarrow$ registration arrow of time one can infer that the spaces $\Phi_+$ and $\Phi_-$ are spaces of Hardy class energy wave functions, respectively, in the upper or lower half of the complex energy plane on the second sheet of the analytically continued S-matrix~\cite{12a}.  As a consequence of the mathematical properties of the Hardy class spaces $\Phi_+$ and $\Phi_-$, the unitary group evolution $U^{\dagger}(t)$, $-\infty < t < \infty$, in $\cal H$ extends to two semigroup evolutions

\begin{equation}
U_+^{\times}(t) = e_+^{-iH^{\times}t} \,\, \mbox{for} \,\, 0 \leq t < \infty \,\, \mbox{in} \,\, \Phi_+^{\times}   \label{eq:gamowdecay}
\end{equation}

\noindent and

\begin{equation}
U_-^{\times}(t) = e_-^{-iH^{\times}t} \,\, \mbox{for} \,\, -\infty < t \leq 0 \,\, \mbox{in} \,\, \Phi_-^{\times}.
\end{equation}

\noindent The asymmetric time evolution of the decaying Gamow vectors is derived from (\ref{eq:gamowdecay}) and (\ref{eq:gamowvector}) for the RHS (\ref{eq:rhsout}):

\begin{equation}
\psi^G(t) = e^{-iH^{\times}t}|E_R-i\frac{\Gamma}{2}^{_-}\rangle = e^{-iE_Rt}e^{-\frac{\Gamma}{2}t}|E_R-i\frac{\Gamma}{2}^{_-}\rangle \,\, \mbox{for} \,\, t \geq 0  \label{eq:gamowtimeev2}
\end{equation}

\noindent This result is identical to the time evolution (\ref{eq:gamowtimeev}), which was conjectured using phenomenological arguments.  Specifically, the time evolution is only defined for $t \geq 0$ in both the phenomenological (\ref{eq:gamowtimeev}) and theoretical (\ref{eq:gamowtimeev2}) results.

There are other Gamow vectors $\tilde{\psi}^G = |E_R+i\frac{\Gamma}{2}^+\rangle \in \Phi_-^{\times}$ which have an asymmetric time evolution under the other semigroup $e_-^{-iH^{\times}t}$ with $t \leq 0$.

\begin{equation}
\tilde{\psi}^G(t) = e^{-iH^{\times}t}|E_R+i\frac{\Gamma}{2}^+\rangle = e^{-iE_Rt}e^{\frac{\Gamma}{2}t}|E_R+i\frac{\Gamma}{2}^+\rangle \,\,\, \mbox{for} \,\,\, t \leq 0
\end{equation}

\noindent The Gamow vectors of the RHS theory have the following features.

\noindent (1) They are derived as functionals over the space of states from the resonance pole term at $z_R = E_R-i\frac{\Gamma}{2}$ (and $z_R^* = E_R+i\frac{\Gamma}{2}$) in the second sheet of the analytically continued S-matrix~\cite{12b}.

\noindent (2) They have a Breit-Wigner energy distribution~\cite{12b}

\begin{equation}
\langle^-E|\psi^G\rangle = i\sqrt{\frac{\Gamma}{2\pi}} \frac{1}{E-(E_R-i\frac{\Gamma}{2})}, \,\,\,\, -\infty_{II} < E < \infty
\end{equation}

\noindent where $-\infty_{II}$ refers to the second sheet.

\noindent (3) The decay probability ${\cal P}(t) = Tr(\Lambda|\psi^G\rangle\langle\psi^G|)$ of $\psi^G(t)$, $t \geq 0$, into the final non-interacting decay products $\Lambda$ can be calculated as a function of time.  From this decay probability, the decay rate is obtained by differentiation $R(t) = \frac{d{\cal P}(t)}{dt}$.  This leads to an exact Golden rule (with the natural linewidth given by a Breit-Wigner) which in the Born approximation becomes Fermi's second Golden rule (of Dirac)~\cite{13}.

\noindent (4) The Gamow vectors $\psi_i^G$ are members of a ``complex'' basis vector expansion~\cite{12b}.  In place of the well known Dirac basis vector expansion (Nuclear Spectral Theorem of the RHS)

\begin{equation}
\phi^+ = \sum_{n} |E_n)(E_n|\phi^+) + \int_{0}^{+\infty} dE|E^+\rangle \langle^+E|\phi^+\rangle
\end{equation}

\noindent (where the discrete sum is over bound states, which we will henceforth ignore), every prepared state vector $\phi^+ \in \Phi_-$ can be expanded as

\begin{equation}
\phi^+ = \sum_{i=1}^{N}|\psi_{i}^{G}\rangle\langle\psi_{i}^{G}|\phi^+\rangle + \int_{0}^{-\infty_{II}}dE|E^+\rangle\langle^+E|\phi^+\rangle  \label{eq:comexp}
\end{equation}

\noindent where $-\infty_{II}$ indicates that the integration along the negative real axis or other contours is on the second Riemann sheet of the S-matrix.  $N$ is the number of resonances in the system (partial wave), each one occurring at the pole position $z_{R_i} = E_{R_i} - i\frac{\Gamma_i}{2}$.  This allows us to mathematically isolate the exponentially decaying states $\psi_i^G$.

The complex basis vector expansion (\ref{eq:comexp}) is rigorous, and one can obtain effective phenomenological theories from it.  The Weisskopf-Wigner approximation methods are tantamount to omitting the background integral, i.e., truncating (\ref{eq:comexp}) to

\begin{equation}
\phi^+ = \sum_{i=1}^{N} |\psi_i^G\rangle c_i
\end{equation}

\noindent where $c_i = \langle\psi_i^G|\phi^+\rangle$.  For instance, for the $K_L$--$K_S$ system with $N = 2$, one has~\cite{7}

\begin{equation}
\phi^+ = \psi_{K_S}^G b_{K_S} + \psi_{K_L}^G b_{K_L}
\end{equation}

\noindent The finite dimensional effective theories which proved so successful as a description of resonance and decay phenomena emerge as a truncation of the complex basis vector expansion of the exact RHS theory.

\section{Summary}

There is a mathematical theory that describes time symmetric as well as time asymmetric quantum physics.  It is an extension of the Rigged Hilbert Space (RHS) formulation of quantum mechanics, which gave a mathematical justification to Dirac's kets and continuous basis vector expansion (circa 1965).

To incorporate causality, this formulation distinguishes meticulously between states and observables by using two different RHS's $\Phi_{\mp} \subset {\cal H} \subset \Phi_{\mp}^{\times}$ of Hardy class functions with complementary analyticity properties.  The dual spaces contain the Dirac kets $|E^{\pm}\rangle \in \Phi_{\mp}^{\times}$ and Gamow vectors $|E_R \pm i\frac{\Gamma}{2}^{\pm}\rangle \in \Phi_{\mp}^{\times}$.  The Gamow vectors have the required properties of resonance states, in particular the enumerated properties (1)--(4) above.  The dual spaces also contain higher order Gamow vectors and higher order Gamow states with exponential time evolution corresponding to higher order S-matrix poles~\cite{14}.  Their Hamiltonian $H^{\times}$ is given on the resonance subspace of $\Phi_+^{\times}$ by a finite dimensional (non-diagonalizable) Jordan block, as expected from the finite dimensional models with non-diagonalizable Hamiltonians~\cite{9}.

The RHS formulation is the mathematical theory of which the Weisskopf-Wigner approximation methods, the Lee-Oehme-Yang theory and all finite dimensional models with complex effective Hamiltonian matrices~\cite{6,7,8,9} can be considered approximations.  Fermi's Golden Rule for the initial decay rate is the Born approximation at $t=0$ of an exact Golden Rule for the decay rate $\dot{{\cal P}}(t)$ which is obtained from the decay probability ${\cal P}(t)$ by differentiation.  

The most surprising feature of the RHS theory is the semigroup time evolution of the Gamow vectors (\ref{eq:gamowtimeev2}) which is a manifestation of a fundamental time asymmetry in quantum mechanics.

\end{document}